\documentstyle[aps,pre,eqsecnum,multicol,epsf]{revtex}

\def\beginwide{
        \end{multicols} \vspace*{-0.5cm} \noindent
        \rule{3.5in}{.1mm}\rule{.1mm}{5mm} \widetext \medskip }
\def\beginwidetop{
        \end{multicols} \vspace*{-0.5cm} \noindent
        \widetext \medskip }
\def\endwide{
        \hspace*{3.5in}~\rule[-5mm]{.1mm}{5mm}\rule{3.5in}{.1mm}
        \begin{multicols}{2} \vspace*{-1.0cm} \noindent }
\def\endwidebottom{
        \begin{multicols}{2} \vspace*{-1.0cm} \noindent }

\begin{document}
\bibliographystyle{unsrt}
\draft

\title{Propagation of an electromagnetic pulse through a waveguide
with a barrier: A time domain solution within classical 
electrodynamics}

\author{Thorsten Emig}

\address{Institut f{\"u}r Theoretische Physik, Universit{\"a}t zu 
K{\"o}ln, D-50923 K{\"o}ln, Germany}

\date{November 1, 1996}
\maketitle

\begin{abstract}
  An electromagnetic truncated Gaussian pulse propagates through a
  waveguide with piecewise different dielectric constants. The
  waveguide contains a barrier, namely a region of a lower dielectric
  constant compared to the neighboring regions. This set-up yields a
  purely imaginary wave vector in the region of the barrier
  ('electromagnetic tunneling').  We exactly calculate the
  time-dependent Green's function for a slightly simplified dispersion
  relation. In order to observe the plain tunneling effect we neglect
  the distortions caused by the wave guide in obtaining the
  transmitted pulse.  The wave front of the pulse travels with the
  vacuum speed of light. Nevertheless, behind the barrier, the maximum
  of the transmitted pulse turns up at an earlier time than in the
  case without an barrier. This effect will be explained in terms of
  the energy flow across the barrier. The solutions obtained reproduce
  the shape of the pulses measured in the tunneling experiments of
  Enders and Nimtz \cite{End1,End2,End3,End4,Nim1}.
\end{abstract}

\pacs{PACS numbers: 03.50.De, 73.40.Gk, 03.40.Kf}

\begin{multicols}{2}
\narrowtext
\section{INTRODUCTION}
\begin{figure}[h]
\begin{center}
\leavevmode
\epsfxsize=1.0\linewidth
\epsfbox{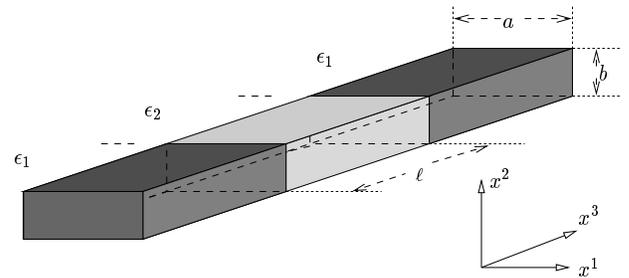}
\caption{The set-up of the considered waveguide.}
\label{fig1}
\end{center}
\end{figure}

Tunneling is, like interference, a characteristic property of
waves. This effect occurs both in non-relativistic quantum mechanics
and in electrodynamics. Although the time dependent differential
equations of both theories are fundamentally different in their
structure, theoretical calculations yield analogous results for the
traversal time of the maximum of a wave packet's modulus.  These
calculations are based on the stationary phase approximation
\cite{Har,Bos} as well as on a scattering ansatz \cite{Mar}. For
sufficiently long barriers, the time delay is independent of the
thickness and, thus, can correspond to arbitrary large effective
velocities of the maximum of a pulse for crossing the barrier. This
prediction is in agreement with results obtained by Nimtz and Enders
in tunneling experiments with evanescent microwaves in a waveguide
with a frequency below cut-off \cite{End1,End2,End3,End4,Nim1}. In
these experiments the evanescent waveguide region, i.e.\ the barrier,
is realized by an undersized region in between normal sized regions of
a waveguide line. Due to the inhomogeneous cross section of the
waveguide, we couldn't obtain an analytic expression 
for the transmission coefficient.
Therefore, a recent microwave experiment \cite{Bro}
studied a barrier produced by a low-dielectric-constant ($\epsilon_2$)
region which was placed in a rectangular waveguide of the same cross
section, filled with a higher dielectric constant
$\epsilon_1$. This experimental set-up is illustrated in 
Fig.\ \ref{fig1}. The relation between wave number $k$ and frequency
$\omega$ is given by the dispersion formula (with the vacuum 
speed of light set to $c=1$)
\begin{equation}
k=\sqrt{\epsilon}\sqrt{\omega^2-\omega_{\rm c}^2/\epsilon}
\label{dispform}
\end{equation}
where $\omega_{\rm c}$ represents the cut-off frequency of the empty 
waveguide and $\epsilon$ is the variable dielectric constant \cite{Jac}. 
Therefore a wave of frequency
$\omega_0$ with $\omega_{\rm c}^2/\epsilon_1 < \omega_0^2 <
\omega_{\rm c}^2/\epsilon_2$ possesses a real wave number $k$ 
outside the barrier, but on entering the tunnel region with a lower
dielectric constant $k$ becomes imaginary, and the wave will spatially
decay. 

In this paper we will consider the electrodynamic tunneling for a
barrier given by a variable dielectric constant.  This set-up is
amenable to a rigorous mathematical description. It was theoretically
investigated first by Martin and Landauer \cite{Mar}. They
concentrated on the tunneling of a Gaussian pulse with a narrow
frequency range. Using a scattering ansatz they showed that the time
delay of the center of mass depends only on the frequency derivative
of the phase of the transmission coefficient. For sufficiently long
barriers, this delay becomes independent of thickness and thus
corresponds to an arbitrary large effective velocity of the center of
mass for crossing the tunnel region, which is known as the Hartman
effect \cite{Har} and has been experimentally verified first 
by Enders and Nimtz \cite{End2}.
We want to examine how this effect is related to causality.  To carry
out this goal, the fundamental solution of the given set-up, the
retarded Green's function, will be constructed analytically assuming
causality. This means that this function vanishes outside the past
light cone, i.e. the wave front travels with the vacuum speed of 
light. With the aid of this solution we will give an analytic 
expression for the entire transmitted pulse, which has the 
observed superluminal property of the center of mass. In addition 
this allows us to determine the deformation of the pulse caused by 
crossing the barrier. It is assumed that the initial pulse is given 
by a truncated Gaussian wave packet located only to the 
left of the barrier.

\section{GREEN'S FUNCTION}
\subsection{The Model}

The Green's function will be obtained using a Laplace transform.
The structure of the dipersion formula (\ref{dispform}) prevents an
analytic inversion of this transformation because of the different
coefficients in front of $\omega^2$ outside and inside the barrier,
respectively. Therefore we consider in this paper a simplified model
for the electromagnetic tunneling effect with dispersion formulas
given by
\begin{equation}
k=\sqrt{\omega^2-m_1^2}, \quad \kappa=\sqrt{\omega^2-m_2^2}
\label{simpdisp}
\end{equation}
with $\sqrt{\epsilon}$ coefficients dropped and cut-off frequencies
$m_1=\omega_{\rm c}/\sqrt{\epsilon_1}$ and $m_2=\omega_{\rm c}/
\sqrt{\epsilon_2}$ outside and inside the barrier, respectively. The
transmission coefficient of the barrier does not change qualitatively
under this simplification. The model given by these dispersion
formulas, together with the Maxwell equations, still represents an
electrodynamic case of tunneling. Within this model the velocity
of a wave front $\lim_{k\to\infty} \omega(k)/k = 1$ is always given
by the vacuum speed of light and it is assumed that the dielectric
medium influences only the cut-off frequencies themselves. 
A similar model consisting of a
classical scalar field which satisfies the one-dimensional
relativistic Klein-Gordon equation with a rectangular potential
barrier was investigated by Deutch and Low \cite{Deu}. Under the
condition that the tunneling amplitude is very small, they found an
approximate solution given by a Gaussian wave packet, which turns up
on the right hand side of the barrier as if its maximum took zero time
to cross the barrier. Our goal here is to find an exact solution of
the Maxwell equations yielding exact values for the tunneling time of
the maximum of a Gaussian wave packet, which can be compared to the
approximate results given by Martin and Landauer. Furthermore,
taking truncated wave packets of variable variance, the influence of
these attributes on the tunneling time will be studied.

Suppose that an electromagnetic pulse has been generated in the region
to the left of the barrier by an appropriate current which has
vanished already before the wave front of the pulse reaches the left
end of the barrier. The propagation of this pulse is then determined
by the dispersion formulas together with the Maxwell equations with
vanishing charge current. One can obtain the propagated field behind
the barrier by solving these equations with the pulse located in front
of the barrier as initial condition.  This can be done with the aid of
the retarded Green's function. In the case of the electromagnetic
field, this fundamental solution can be written as an antisymmetric 
tensor $G_{\bar{x}}{}_{\alpha\beta}$, $(\alpha,\beta=0,\ldots,3)$, 
satisfying the four dimensional wave equation with the Dirac 
distribution at the fixed space-time position $\bar{x}$ as inhomogeneity
\begin{equation}
\Box G_{\bar{x}}=-\delta_{\bar{x}},
\label{waveeq}
\end{equation}
where $\Box=\partial_0^2-\partial_1^2-\partial_2^2-\partial_3^2$ 
is the Laplace-Beltrami-operator of the Minkowskian space
time \cite{Thi}. Beyond that, $G_{\bar{x}}$ must allow for the
following boundary conditions: First, it has to vanish outside the
past light cone at $\bar{x}$, second its six components have to
jump in the respective correct manner at the two boundaries between the
propagating and the evanescent region and must take into account the
metallic boundary conditions on the surface of the waveguide. In the
next section, we will give a solution of this boundary value problem
where the axial magnetic component $G_{\bar{x}}{}_{12}$ of Green's
function shall serve as example. Given the relevant field components
in front of the barrier at the time $x^0=0$, the expression
\begin{equation}
\label{b3int}
B_3(\bar{x})=\int\limits_N \left[ 
G_{\bar{x}}{}_{12,1}E_2 
-G_{\bar{x}}{}_{12,2}E_1 
-G_{\bar{x}}{}_{12,0}B_3 \right] dx^1 dx^2 dx^3
\end{equation}
yields the time evolution of the $B_3$ component of the
magnetic field behind the barrier $\bar{x}^3>\ell/2$ in terms
of partial derivatives $G_{\bar{x}}{}_{\alpha\beta,\gamma}$ of Green's
function.  The three-dimensional domain $N=\{x\in{\bf R}^4 | 0\le x^1
\le a, 0\le x^2 \le b, x^0=0 \}$ of integration is the interior of the
whole waveguide line at time $x^0=0$.

\subsection{Analysis}

We solve the time dependent wave equation (\ref{waveeq}) by a Laplace
transform relative to the time $x^0$ for a fixed $\bar{x}$ with the
property $\bar{x}^0>0$. The Laplace transform yields, after inversion, a
function in the time domain which vanishes {\em until} a certain time
is reached. Therefore this method is only suitable to get the advanced
Green's function. The retarded Green's function, though, can then be
obtained by a simple time reflection.

The transformed components of Green's function are defined by
\begin{equation}
\tilde{G}_{\bar{x}}{}_{\alpha\beta}(\omega)={\cal
  L}[G_{\bar{x}}{}_{\alpha\beta}]=\int_0^{\infty}e^{i\omega
  x^0}G_{\bar{x}}{}_{\alpha\beta}dx^0
\end{equation}
where ${\cal L}$ denotes the Laplace transform with $\omega$ in 
the upper half plane of complex numbers.

For a waveguide with perfectly conducting walls given by the surface
$S$, the boundary condition for the electromagnetic field reads
\cite{Jac}
\begin{equation}
(\vec{n}\times\vec{E})_{|S}=0,\quad (\vec{n}\cdot\vec{B})_{|S}=0
\end{equation}
where $\vec{n}$ is the vector normal to the surface $S$.  This
condition is satisfied if one uses a Green's function which is
invariant relative to the reflections at the four walls of the
rectangular waveguide. This invariant Green's function can be
decomposed into the characteristic modes of the waveguide. The result
for the Laplace transformed components can be written as
\begin{equation}
\tilde{G}_{\bar{x}}{}_{\alpha\beta}=\frac{1}{2\pi
  ab}\sum_{\iota,\eta=-\infty}^{\infty}\!\! f_{\alpha\beta} 
e^{-i(x^1\pi/a \iota+x^2\pi/b \eta)}e^{i\omega\bar{x}^0}
\tilde{g}_{\alpha\beta}(\bar{x}^3,x^3)
\label{ansatz1}
\end{equation}
introducing the antisymmetric tensor
\begin{equation}
f_{\alpha\beta}=
\left(\begin{array}{cccc}
0 & ic_1s_2 & is_1c_2 & -s_1s_2\\
-ic_1s_2 & 0 & c_1c_2 & ic_1s_2\\
-is_1c_2 & -c_1c_2 & 0 & is_1c_2\\
s_1s_2 & -ic_1s_2 & -is_1c_2 & 0\\ \end{array}\right)
\end{equation}
with
\begin{eqnarray}
c_1:=\cos(\bar{x}^1\iota\pi/a),\quad s_1:=\sin(\bar{x}^1\iota\pi/a),\\
c_2:=\cos(\bar{x}^2\eta\pi/b),\quad s_2:=\sin(\bar{x}^2\eta\pi/b).
\end{eqnarray}
This tensor describes the $x^1$- and $x^2$-dependence of the modes
represented by its respective pairs of numbers $(\iota,\eta)$. Here we
have introduced a one dimensional Green's function
$\tilde{g}_{\alpha\beta}(\bar{x}^3,x^3)$ to account for the dependence
on the axial coordinate $x^3$. Only this function contains the
dependence on the frequency and will therefore define the time
evolution of the field.  Applying the transformed wave equation
(\ref{waveeq}) to Eq.\ (\ref{ansatz1}), we find that the one
dimensional Green's function has to be a solution of the ordinary
differential equation given by
\begin{equation}
\left[\frac{d^2}{d(x^3)^2}+\omega^2-m^2(x^3)\right]
\tilde{g}_{\alpha\beta}(\bar{x}^3,x^3)=\delta(x^3-\bar{x}^3)
\label{1dgreen}
\end{equation}
with the jumping cut-off frequency determined by the simplified
dispersion formulas (\ref{simpdisp}), i.e.
\begin{equation}
m(x^3)=\left\{\begin{array}{ll}
m_1=\omega_{\rm c}/\sqrt{\epsilon_1}, & |x^3|\ge \ell/2\\
m_2=\omega_{\rm c}/\sqrt{\epsilon_2}, & |x^3|<\ell/2,
\end{array} \right.
\end{equation}
with $\omega_{\rm c}=\pi\sqrt{\iota^2/a^2+\eta^2/b^2}$ for a
rectangular waveguide. Notice that the cut-off frequencies $m_1$ and
$m_2$ change with the type of mode. Thus the one dimensional Green's
function also depends on the numbers $(\iota,\eta)$. In 
Eq.\ (\ref{1dgreen}) no boundary terms at zero time emerge from the 
Laplace transform of the second time derivative due to the vanishing 
of the advanced Green's function for $x^0<\bar{x}^0$ and the 
property $\bar{x}^0>0$.

The one dimensional Green's function for
(\ref{1dgreen}) can be constructed from two linear independent 
solutions of the corresponding homogeneous equation. These can be 
chosen as a plane wave traveling from the left-hand side and the 
right-hand side, respectively, toward the barrier, i.e.
\begin{eqnarray}
\label{funda1}
\phi^{(1)}_{\alpha\beta}(x^3)&=&\left\{\begin{array}{ll}
\epsilon_{\alpha\beta}e^{ikx^3}+ R_{\alpha\beta}e^{-ikx^3},
& x^3\le -\ell/2, \\
A_{\alpha\beta}e^{i\kappa x^3}+B_{\alpha\beta}e^{-i\kappa 
x^3},
& |x^3|<\ell/2, \\
T_{\alpha\beta}e^{ikx^3},
& x^3\ge \ell/2, \end{array}\right.
\end{eqnarray}
and
\begin{eqnarray}
\phi^{(2)}_{\alpha\beta}(x^3)&=&\left\{ \begin{array}{ll}
T_{\alpha\beta}e^{-ikx^3},
& x^3\le -\ell/2, \\
A_{\alpha\beta}e^{-i\kappa x^3}+B_{\alpha\beta}e^{i\kappa 
x^3},
& |x^3|<\ell/2, \\
\epsilon_{\alpha\beta}e^{-ikx^3}+R_{\alpha\beta}e^{ikx^3},
& x^3\ge \ell/2. \end{array}\right.
\label{funda2}
\end{eqnarray}
with the total antisymmetric tensor $\epsilon_{\alpha\beta}$.
The wave numbers $k$ and $\kappa$ are given by the dispersion formulas
(\ref{simpdisp}). Due to the upper half plane analyticity of a Laplace
transformed function, the square roots in these equations have to be
chosen analytic in the upper $\omega$ half plane. Hence the wave
numbers always have a positive imaginary part. The respective
coefficients $R_{\alpha\beta}$, $A_{\alpha\beta}$, $B_{\alpha\beta}$
and $T_{\alpha\beta}$ are uniquely determined by the boundary
conditions at the two planes defining the evanescent region between
$x^3=-\ell/2$ and $x^3=\ell/2$. The coefficients of both solutions are
assumed to be equal because of the symmetry of the barrier relative to
$x^3=0$. Taking into account that the imaginary part of $k$ is
positive, the asymptotic behavior of the two solutions looks like
\begin{equation}
\lim_{x^3\to -\infty} \phi^{(2)}_{\alpha\beta}(x^3)=0, \quad  
\lim_{x^3\to\infty}\phi^{(1)}_{\alpha\beta}(x^3)=0.
\end{equation}
Therefore we can construct Green's function in the upper $\omega$
half plane as
\begin{eqnarray}
\tilde{g}_{\alpha\beta}(\bar{x}^3,x^3)=\frac{1}{W_{\alpha\beta}}
\cdot\left\{\begin{array}{ll}
\phi^{(1)}_{\alpha\beta}( 
\bar{x}^3)\phi^{(2)}_{\alpha\beta}(x^3),
& \bar{x}^3\ge x^3\\
\phi^{(2)}_{\alpha\beta}( 
\bar{x}^3)\phi^{(1)}_{\alpha\beta}(x^3),
& \bar{x}^3\le x^3\end{array}\right.
\label{green1fct}
\end{eqnarray}
with the Wronskian $W_{\alpha\beta}$ of $\phi^{(1)}_{\alpha\beta}$ and
$\phi^{(2)}_{\alpha\beta}$ defined by
\begin{equation}
W_{\alpha\beta}=\phi^{(2)}_{\alpha\beta} 
\frac{d\phi^{(1)}_{\alpha\beta}}{dx^3}-
\phi^{(1)}_{\alpha\beta} 
\frac{d\phi^{(2)}_{\alpha\beta}}{dx^3}.
\label{wronskian}
\end{equation}

The case $x^3<\bar{x}^3$ with $x^3<-\ell/2$ and $\bar{x}^3>\ell/2$ is
the interesting one for the investigation of the tunneling time
problem for wave packets. In this case Green's function is suitable to
calculate the field behind the barrier for a pulse assumed to be
located initially only in the front of the barrier. In this case the
Green's function reads
\begin{equation}
\tilde{g}_{\alpha\beta}(\bar{x}^3,x^3)=\frac{1}{2ik} 
T_{\alpha\beta} e^{ik(\bar{x}^3-x^3)}.
\label{green-icase}
\end{equation}

We now must consider the respective matching conditions for the
components of the electromagnetic field to evaluate the coefficients
of the two solutions. For simplicity let us only look at the
calculations for the component $B_3$. This component has to be
continuous at $x^3=-\ell/2$ and $x^3=\ell/2$. Furthermore we need the
behavior of its partial derivative perpendicular to the
boundary-planes of the barrier.  The transverse components
$B_1$ and $B_2$ are continuous at any position
of the two boundary-planes. Hence its partial derivatives parallel to
the boundary-planes, i.e. the $x^1$- and $x^2$-derivatives, also have
to be continuous at these planes. The
equation div$\vec{B}=0$ shows that the $x^3$-derivative of the
component $B_3$ is continuous at the boundary-planes, too. These four
continuity conditions together yield the coefficients of the
corresponding functions $\phi^{(1)}_{12}$ and $\phi^{(2)}_{12}$:
\begin{eqnarray}
R_{12}&=&D^{-1}e^{-ik\ell}\left(1-e^{2i\kappa 
\ell}\right)\left(k^2-
\kappa^2\right)\\
A_{12}&=&D^{-1}2k(k+\kappa)e^{i(
\kappa-k)\ell/2}\\
B_{12}&=&-D^{-1}2k(k-
\kappa)e^{i(3\kappa-k)\ell/2}\\
T_{12}&=&D^{-1}4k\kappa e^{i(\kappa-
k)\ell}
\label{tc}
\end{eqnarray}
with the common denominator
\begin{equation}
D=(k+\kappa)^2-(k-\kappa)^2 e^{2i\kappa \ell}.
\end{equation}
Due to the absence of upper half plane zeros of $D$ the coefficients
are analytic in this half plane.

We find the one dimensional retarded Green's function in the time
domain after a time reflection. The inverse Laplace transform yields,
for $t:=\bar{x}^0-x^0\ge 0$,
\begin{equation}
g_{\alpha\beta}(t,\bar{x}^3,x^3)=
\frac{1}{2\pi} \int_{-\infty+is}^{\infty+is} e^{-
i\omega t} \tilde{g}_{\alpha\beta}(\bar{x}^3,x^3) d\omega
\label{invers}
\end{equation}
with $\tilde{g}_{\alpha\beta}(\bar{x}^3,x^3)$ given by Eq.\
(\ref{green-icase}). 

In the case of the component $B_3$ we now have to consider the
transmission coefficient $T_{12}$ given in (\ref{tc}). Note that the
expression $k+\kappa$ never vanishes in the upper $\omega$ half
plane. Thus, expanding the right-hand side of (\ref{tc}) with
$1/(k+\kappa)^2$, we find
\begin{equation}
T_{12}=\frac{4k\kappa}{(k+\kappa)^2} e^{i(\kappa-
k)\ell}\frac{1}{1-\left(\frac{k-\kappa}{k+\kappa}\right)^2 
e^{2i\kappa \ell}}.
\end{equation}
The term $(k-\kappa)/(k+\kappa)$ is bounded in the upper $\omega$ half
plane by $1$ and the imaginary part of $\kappa$ is always positive in
this half plane. Thus the transmission coefficient can be written as
a geometric series
\begin{equation}
T_{12}=4k\kappa e^{-ik\ell}\sum_{\nu=0}^{\infty} \frac{(k-
\kappa)^{2\nu}}{(k+\kappa)^{2\nu+2}} e^{i\kappa \ell 
(2\nu+1)}.
\end{equation}

At this stage we need the simplified dispersion formulas
(\ref{simpdisp}) for an analytic inversion of the Laplace
transform. Expanding each term of the series with $(k+\kappa)^{2\nu}$
cancels the $\omega^2$ terms in the numerator. Thus the transformed
Green's function (\ref{green-icase}) becomes
\begin{eqnarray}
\tilde{g}_{12}(\bar{x}^3,x^3)&=&-2i\kappa^2 e^{ik(\bar{x}^3-x^3-\ell)}
\nonumber\\
&&{} \times\sum_{\nu=0}^{\infty}\frac{1}{\kappa} \frac{(m_2^2 - 
m_1^2)^{2\nu}}{(k+\kappa)^{4\nu+2}} e^{i\kappa\ell(2\nu+1)}.
\end{eqnarray}
For our further calculations it is useful to write this as
\begin{equation}
\tilde{g}_{12}(\bar{x}^3,x^3)=\frac{2}{m_0^2}\sum_{\nu=0}^ 
{\infty} \tilde{f}_{\nu}\left(\sqrt{\omega^2-m_1^2}\right)
\label{gsh}
\end{equation}
with the new function $\tilde{f}_{\nu}$ given by
\begin{eqnarray}
\label{fdef}
\tilde{f}_{\nu}(\omega)&=&(\omega^2-m_0^2)e^{i\omega(\bar{x}^3-
x^3-\ell)} \\ \nonumber
&&{}\times \frac{-i m_0^{4\nu+2} 
e^{i(2\nu+1)\ell\sqrt{\omega^2-m_0^2}}} {\sqrt{\omega^2-
m_0^2}\left(\omega+\sqrt{\omega^2-m_0^2}\right)^{4\nu+2}}
\end{eqnarray}
and an effective cut-off frequency $m_0=\sqrt{m_2^2-m_1^2}$.
Disregarding the factor ahead of the fraction in
(\ref{fdef}) for a moment, the function
$\tilde{f}_{\nu}$ corresponds to the time domain function \cite{Bat}
\beginwide
\begin{equation}
 {\cal L}^{-1}\left[ \tilde{f}_{\nu}(\omega)\frac{e^{-
i\omega(\bar{x}^3-x^3-\ell)}}{\omega^2-
m_0^2}\right]
=\Theta\left(t-
(2\nu+1)\ell\right)\left(\frac{t-
(2\nu+1)\ell}{t+(2\nu+1)\ell}\right)^{2\nu+1} 
J_{4\nu+2}\left(m_0\sqrt{t^2-(2\nu+1)^2 \ell^2}\right),
\end{equation}
with the step function $\Theta(x)$ and the $\nu$th Bessel function 
$J_\nu(x)$.

The additional exponential factor only causes a time translation. By
performing the inverse Laplace transformation ${\cal L}^{-1}$ one gets
for $q_{\nu}(t,z)={\cal
L}^{-1}[\tilde{f}_{\nu}(\omega)(\omega^2-m_0^2)^{-1}]$, with the
abbreviation $z=\bar{x}^3-x^3>0$, the solution
\begin{equation}
q_{\nu}(t,z)=\Theta(t-z-2\nu\ell)\left(\frac{t-z+\ell-
(2\nu+1)\ell}{t-z+\ell 
+(2\nu+1)\ell}\right)^{2\nu+1}
J_{4\nu+2}\left(m_0\sqrt{(t-z+\ell)^2-(2\nu+1)^2 
\ell^2}\right).
\label{helpf1}
\end{equation}

Because of the remaining factor $\omega^2-m_0^2$ the second time
derivative of $q_{\nu}(t,z)$ enters the function $f_{\nu}(t,z)={\cal
  L}^{-1}[\tilde{f}_{\nu}(\omega)]$. Due to the vanishing of the
function $q_{\nu}(0,z)$ for the values $z+2\nu\ell>0$ no boundary
terms arise from the Laplace transform of the second time derivative
$\ddot{q}_{\nu}(t,z)$. Thus we have
\begin{equation}
f_{\nu}(t,z)=-m_0^2 q_{\nu}(t,z)-\ddot{q}_{\nu}(t,z).
\end{equation}
Now the inverse transform of the series in (\ref{gsh}) yields for 
its terms, defined by $h_{\nu}(t,z)={\cal L}^{-1}[\tilde{f}_{\nu}
(\sqrt{\omega^2-m_1^2})]$, the expression \cite{Bat}
\begin{equation}
h_{\nu}(t,z)=f_{\nu}(t,z)-m_1\int\limits_0^t 
f_{\nu}\left(\sqrt{t^2-u^2},z\right)J_1(m_1 u) du.
\label{helpf2}
\end{equation}
Therefore we obtain for the integral (\ref{invers}) the series
expansion
\begin{eqnarray}
g_{12}(t,z) &=& \frac{2}{m_0^2}\sum_{\nu=0}^{\infty} 
h_{\nu}(t,z) \nonumber \\ 
&=& -2\sum_{\nu=0}^{\infty} 
\left[q_{\nu}(t,z)+\frac{1}{m_0^2} \ddot{q}_{\nu}(t,z)\right.
\nonumber\\
&& {}\left. -m_1\int_0^t\left\{q_{\nu}\left(\sqrt{t^2-
u^2},z\right)+\frac{1}{m_0^2} \ddot{q}_{\nu}\left(\sqrt{t^2-
u^2},z\right)\right\} J_1(m_1 u) du \right]. 
\end{eqnarray}
\endwide

The structure of this Green's function can be explained
phenomenologically by looking at Eq.\ (\ref{helpf2}).  Even the leading
term $q_0$ of the series, and thus $h_0$, jumps from zero to a finite
value at the boundary of the past light cone. This property of
Green's function guarantees the causal propagation of every pulse. In
opposition to the free space here the support of Green's function is
not only the boundary of the light cone but its full interior. There
are two reasons for this feature. First, consider the first term in
Eq.\ (\ref{helpf2}) and accordingly the terms of the series $q_{\nu}$
given by Eq.\ (\ref{helpf1}). These terms contribute to the series
only when $t>z+2\nu\ell$. Thus the $\nu$th term of the series
represents the part of the field that leaves the barrier on the right
hand side after $2\nu$ reflections at its boundaries. For the second
reason take a look at the second term of Eq.\ (\ref{helpf2}). This
term has nothing to do with the barrier itself but arises from the
boundary conditions of the wave guide. Notice that the field is
reflected there and back from the metallic boundaries while
propagating through the waveguide. This echo effect is described by
the integral of the term in question. This integral represents a
distortion in which all excitations that are noticeable at a given
position in the time interval $[0,t]$ take part.

\section{TUNNELING OF WAVE PACKETS}
\subsection{Analysis}

Now we will solve the Maxwell equations for a given pulse using
Green's function determined in the last section. Before starting with
this calculation we want to simplify Green's function. It has been 
noticed above that a distortion integral like that of
Eq.\ (\ref{helpf2}) always arises in the case of guided
waves. However, the tunneling effect itself is described completely by
the functions $f_{\nu}(t,z)$ in the first term of Eq.\ (\ref{helpf2}).
We are interested in the undistorted delay induced by the tunneling
effect only.
Therefore, in the following calculations we consider only this term of
Green's function, i.e. we set
\begin{equation}
g_{12}(t,z)=-2\sum_{\nu=0}^{\infty} 
\left\{q_{\nu}(t,z)+\frac{1}{m_0^2} \ddot{q}_{\nu}(t,z)\right\}
\label{simpgreen}
\end{equation}
to determine the influence of the barrier itself on the transmitted
pulse.  With this simplification, the studied system has been mapped
onto the model investigated by Deutch and Low \cite{Deu}.  Their cut
off frequency corresponds to our effective cut-off frequency
$m_0=\sqrt{m_2^2-m_1^2}$. Notice that, for comparing with experimental
results, the contribution of the distortion term to the total
transmitted pulse decreases with an increasing difference between the
lower cut-off frequency $m_1$ and the central frequency $\omega_0$ of
the pulse.

Using now Eq.\ (\ref{b3int}), (\ref{ansatz1}) with the simplified form
(\ref{simpgreen}) of Green's function with $x^0=0$, we obtain the
component $B_3$ behind the barrier for an arbitrary pulse started in
front of the tunneling region.  Our calculations will be carried out
with a pulse given by a truncated Gaussian $H_{10}$-mode centered at
some position $x^3=s<-\ell/2$ in front of the barrier and with central
frequency $\omega_0$ corresponding to the wavenumber
$k_0=\sqrt{\omega_0^2-m_1^2}$. In this case the relevant and non
vanishing field components \cite{Jac} are the real parts of
\begin{eqnarray}
\label{initial}
E_2(x)&=&i\omega_0 \frac{a}{\pi} \sin(x^1 \pi/a) 
\,\varphi(x^3)\\
B_3(x)&=&\cos(x^1 \pi/a)\,\varphi(x^3)
\end{eqnarray}
with the Gaussian envelope
\begin{equation}
\varphi(x^3)= \Theta(-x^3+s+\gamma)\,e^{ik_0 x^3} 
e^{-(x^3-s)^2/\sigma^2}.
\label{gaussfkt}
\end{equation}
The distance between the maximum and the wavefront of the packet is
given by the parameter $\gamma$ with the property
$0<\gamma<-s-\ell/2$. The upper boundary of $\gamma$ comes from the
condition that the wavefront of the initial pulse has to be in front
of the barrier at $x^0=0$. This is necessary because the barrier
causes deformations of the pulse which are initially unknown.

\begin{figure}[h]
\begin{center}
\leavevmode
\epsfxsize=1.0\linewidth
\epsfbox{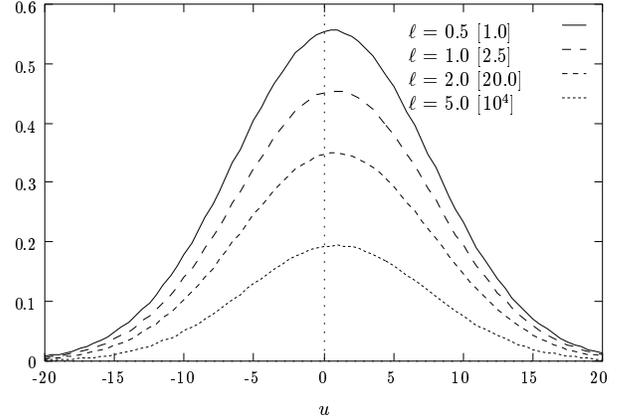}
\caption{Graphs of the envelope of the component 
$B_3$ of the magnetic field for Gaussian wave packets 
transmitted below cut-off across 
barriers of different thickness $\ell$ as a function of the 
relative coordinate $u$. The cut-off frequencies are $m_1=1$ and 
$m_2=4$. In all cases $\omega_0=3.2$ and $\sigma=10$. The graphs 
are scaled by the factors in the brackets.}
\label{fig2}
\end{center}
\end{figure}

With this initial pulse one can carry out the integration in 
Eq.\ (\ref{b3int}) using the convolution theorem.
Changing the integration variable to $v=t-\bar{x}^3+\ell$, we find
the final solution for the component $B_3$ of the pulse
behind the barrier. With the relative coordinate $u=\bar{x}^0-
\bar{x}^3+s+\ell$ one gets
\beginwidetop
\begin{equation}
\label{finsol}
B_3(\bar{x})=-\cos(\bar{x}^1\pi/a)\frac{2}{m_0^2} 
\sum_{\nu=0}^{\infty}\left[\cos(k_0(u-
s))\Gamma^{(1)}_{\nu}(u)
+\sin(k_0(u-s))\Gamma^{(2)}_{\nu}(u)\right]
\end{equation}
with
\begin{eqnarray}
\Gamma^{(1,2)}_{\nu}(u)&=&\Theta(u+\gamma-
(2\nu+1)\ell)\int\limits_{(2\nu+1)\ell}^{u+\gamma}\! e^{-(u-
v)^2/\sigma^2} \left( \frac{v-(2\nu+1)\ell}{v+(2\nu+1)\ell}
\right)^{2\nu+1}
\nonumber\\
&&{}\times J_{4\nu+2}\left(m_0\sqrt{v^2-(2\nu+1)^2\ell^2}\right)
\left(\cos(k_0 v)\vartheta^{(1,2)}(u-
v)\mp\sin(k_0v)\vartheta^{(2,1)}(u-v)\right) dv,\nonumber\\
\label{gamma}
\end{eqnarray}
and the abbreviations
\begin{eqnarray}
\vartheta^{(1)}(t) &=& \frac{2}{\sigma^2}t\left( \omega_0^2-
m_2^2+2k_0(k_0+\omega_0)-\frac{4t^2}{\sigma^4}+\frac{6}{\sigma^2} 
\right)\nonumber \\ \\
\vartheta^{(2)}(t) &=& k_0\left( \omega_0^2-m_2^2-
\frac{12t^2}{\sigma^4}+\frac{6}{\sigma^2}\right)
+\omega_0 \left( \omega_0^2-m_2^2-
\frac{4t^2}{\sigma^4}+\frac{2}{\sigma^2}\right).
\end{eqnarray}
\endwide
The integral in Eq.\ (\ref{gamma}) can be evaluated only
numerically. Note that, for a fixed value of $u$, all of the
functions $\Gamma_{\nu}^ {(1,2)}(u)$ with an index $\nu$ larger than
some index $\nu_0$ vanish.  Naturally, the index $\nu_0$ depends on
the value of $u$. So we have to do only a finite number of numerical
integrations to obtain the component $B_3$. The envelope of $B_3$ is
shown in Fig.\ \ref{fig2} for different barrier thicknesses $\ell$ and
fixed central and cut-off frequencies. The distance $\gamma$ between
the maximum and the wave front of the wave packet has to be large
enough to prevent deformations of the transmitted pulse that arise in
the case of a pulse with too large high-frequency components. For this
and the following results $\gamma$ has been chosen to be five times
the initial variance $\sigma_0$ of the pulse.  Due to the trivial
dependence of the solution on the coordinate $x^1$ we have always set
$x^1=0$ corresponding to the boundary of the waveguide.  Because we
have eliminated the echo effect caused by the waveguide itself, the
pulse does not change its shape outside the barrier region. Thus the
graphs in Fig.\ \ref{fig2} correspond to the time evolution of the
wave packet measured by an observer at an arbitrary position behind
the barrier.  To determine the tunneling time of the maximum of the
wave packet, we can use the free propagation of the wave packet
outside the barrier, i.e.\ the fact that it travels there with the
vacuum speed of light without changing its shape. 
Then the maximum of the
packet arrives at the left end of the barrier at $x^0=-s-\ell/2$. Now
let $\tau$ be the value of $u$ at which the envelope of $B_3(u)$ has
its maximum. Considering the definition of the coordinate $u$, one
obtains for the arrival time at the right end of the barrier
$x^0=\tau-s-\ell/2$. Thus the tunneling time is given by $\tau$.
\narrowtext
The graphs in Fig.\ \ref{fig2} show that the transmitted pulses are
also Gaussian-like wave packets, but exponentially damped with growing
barrier thickness. To obtain the tunneling time and the variance of
the packets we have fitted Gaussian wave packets to the graphs of
Fig.\ \ref{fig2}. The resulting values for $\tau$ and the variance
$\sigma$ are listed in Table I. Furthermore, this table shows
values also corresponding to Gaussian-like solutions for other cut-off
frequencies of which the graphs are not shown here. While in the case
of $\ell=0.5$ the tunneling time corresponds to a subluminal average
velocity of the maximum, it increases more and more slowly with
growing barrier thickness. This result agrees with the experimental
observation that for sufficiently long barriers the tunneling time is
independent of thickness \cite{End2}. The variance of the transmitted
wave packets decreases with increasing barrier thickness.

To determine the dependence of the tunneling time on the central
frequency of the initial pulses we have calculated the corresponding
transmitted wave packets for a barrier of fixed thickness $\ell=5$
with $m_1=1$ and $m_2=4$. We have considered Gaussians with variances
both above ($\sigma_0=10$) and below ($\sigma_0=4$) the thickness. The
resulting pulses are also Gaussian-like except for the
low-variance-pulses with $\omega_0 \stackrel{>}{\sim} m_2/2$. The
parameters of the Gaussian solutions are given by 
Table II. The tunneling time increases with the central 
frequency but corresponds always to a superluminal average velocity 
for the maximum of the pulse. The maxima are shifted to higher values 
of $\tau$ with increasing $\omega_0$ because of contributions from
Fourier components above the barrier cut-off. These components are
also responsible for the slightly higher times of the narrow packets
and its distortion at higher central frequencies. The variance of the
transmitted packets decreases with increasing $\omega_0$.

{\small TABLE I. Numerical results for tunneling times and 
variances of wave packets that have crossed barriers of different 
length. Two different pairs of cut-offs are considered with a 
corresponding central frequency in between. In both cases the 
initial variance is $\sigma_0=10$.}

\begin{center}
\begin{tabular*}{8.5cm}{d@{\extracolsep\fill}dddd}
\hline\hline
 $\ell$ & \multicolumn{2}{c}{$m_1=1$, $m_2=4$} &
\multicolumn{2}{c}{$m_1=6.5$, $m_2=9.5$} \\
 & \multicolumn{2}{c}{$\omega_0=3.2$} & \multicolumn{2}{c}
{$\omega_0=8.5$} \\
 & $\tau$ & $\sigma$ & $\tau$ & $\sigma$ \\ \hline
 0.5 & 0.66 & 10.01 & & \\
 1.0 & 0.82 & 9.97 & & \\
 2.0 & 0.86 & 9.85 & & \\
 3.0 & & & 0.48 & 9.82 \\
 5.0 & 0.91 & 9.38 & 0.49 & 9.66 \\ \hline\hline
\end{tabular*}
\end{center}

\begin{figure}[t]
\begin{center}
\leavevmode
\epsfxsize=1.0\linewidth
\epsfbox{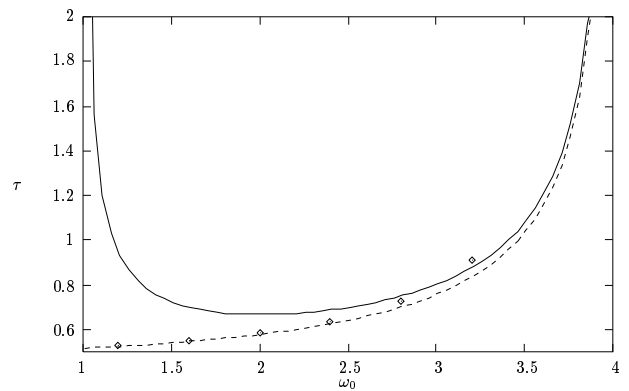}
\caption{Graph of the tunneling time for a wave packet calculated by Martin and
Landauer \protect\cite{Mar} from the frequency derivative of the phase
of the transmission coefficient as a function of the central frequency
of the packet. Solid line: The cut-off frequencies are $m_1=1$,
$m_2=4$ and the barrier thickness is $\ell=5$. The dashed line
corresponds to a vanishing cut-off frequency outside the barrier
($m_1=0$). The dots represent the tunneling time of the maximum of our
solution given by Eq.\ (\ref{finsol}) with $\sigma_0=10$.}
\label{fig3}
\end{center}
\end{figure}

\begin{figure}[htb]
\begin{center}
\leavevmode
\epsfxsize=1.0\linewidth
\epsfbox{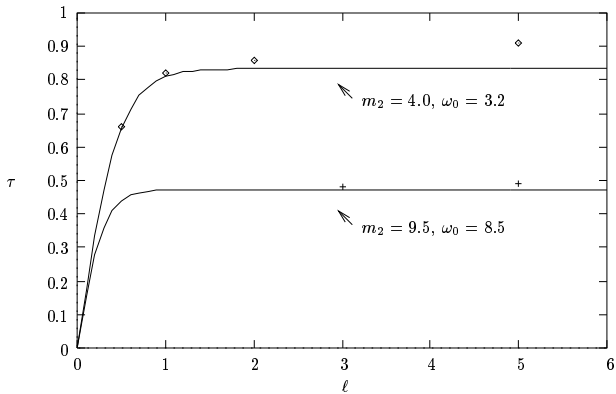}
\caption{Graphs of the tunneling time as given both by \protect\cite{Mar} with
vanishing cut-off frequency $m_1$ outside the barrier (lines) and by
the maximum of the solution (\ref{finsol}) (dots) as a function of the
barrier thickness for two different pairs of cut-off frequencies and a
central frequency in between. The dots correspond to the values given
in Table I.}
\label{fig4}
\end{center}
\end{figure}

Let us now compare the tunneling times obtained from our solutions
with that of Martin and Landauer \cite{Mar}. These authors have
pointed out that the time delay of the center of mass for a pulse
restricted to a wide variance in the time domain depends only on the
frequency derivative of the phase $\alpha$ of the transmission
coefficient, i.e.
\begin{equation}
\tau_{\rm ML}=\frac{d\alpha}{d\omega}+\ell\frac{dk}{d\omega},
\label{MLdelay}
\end{equation}
where the expression has to be evaluated at $\omega=\omega_0$.  Due to
the symmetry of the transmitted pulse the time delay for the center of
mass and the maximum are the same. But the time $\tau_{\rm ML}$ does
not represent the pure time delay of the maximum caused by the barrier
itself: Due to the echo effect of the waveguide which also affects the
transmission coefficient, $\tau_{\rm ML}$ is shifted to higher
values. To account for this fact, we compare the tunneling times of
our echo-free solutions with that of Martin and Landauer by taking in
Eq.\ (\ref{MLdelay}) the phase of the transmission coefficient with
$m_1=0$, corresponding to a free propagation outside the barrier. The
graph of $\tau_{ML}$ and our values of $\tau$ are plotted in 
Figs.\ \ref{fig3} and \ref{fig4} as functions of the central frequency 
and the barrier thickness, respectively, for the parameters considered
above. A difference between the two tunneling times arises only for
higher central frequencies because of the growing contribution of high
Fourier components to the transmitted pulse. Notice that the
approximate result of Martin and Landauer is valid only for pulses
with a narrow frequency range. Furthermore our values for the
tunneling time retains a small dependence on thickness, but
corresponds nevertheless to a superluminal velocity for the maximum of
the pulse. At this moment it should be emphasized again that the wave
front travels always with the vacuum speed of light. 
Below, the possibility of superluminal maxima within an underlying 
causal propagation will be explained in terms of the energy flow 
across the barrier.

{\small TABLE II. Numerical results for tunneling times and 
variances of wave packets with different central frequencies 
between both cut-offs and initial variance $\sigma_0=10$ and 
$\sigma_0=4$, respectively. The barrier thickness is $\ell=5$.}

\begin{center} 
\begin{tabular*}{8.5cm}{d@{\extracolsep\fill}dddd}
\hline\hline
 $\omega_0$ & \multicolumn{2}{c}{$\sigma_0=10$} & 
\multicolumn{2}{c}{$\sigma_0=4$} \\
 & $\tau$ & $\sigma$ & $\tau$ & $\sigma$ \\ \hline
 1.2 & 0.52 & 9.90 & 0.53 & 3.72 \\
 1.6 & 0.55 & 9.87 & 0.57 & 3.61 \\
 2.0 & 0.58 & 9.83 & 0.63 & 3.43 \\
 2.4 & 0.64 & 9.78 & & \\
 2.8 & 0.72 & 9.68 & & \\
 3.2 & 0.91 & 9.38 & & \\ \hline\hline
\end{tabular*}
\end{center}

\subsection{Interpretation}

Apparently, the maxima of the solutions we have given above
cross the barrier with a superluminal velocity. In this section, we
want to explain this property of the transmitted wave packet in terms
of the energy flow across the barrier. In Minkowskian space-time it is
given by the integral curves of the 4-vector field ${\cal T}^{\alpha}$
with the electromagnetic energy density ${\cal T}^0= \frac{1}{2}
(\vec{E}\cdot\vec{D}+\vec{B}\cdot\vec{H})$ and the Poynting vector
${\cal T}^j=(\vec{E}\times\vec{H})_j$ for $j=1,2,3$.

The fact that ${\cal T}^{\alpha}$ is a continuous differentiable
vector field induces an important property of its integral curves in
space-time: Curves with different initial positions do not intersect
each other. This property is not destroyed by the jumping conditions
for the electromagnetic field at the ends of the barrier. To prove
this claim we have to show only that any curve flowing on a boundary
of the barrier from one side has a unique continuation at the other
side of the boundary. This means that there are no branching points
for the curves at the boundary. Due to the continuity of the component
${\cal T}^3$ at the boundary these points could only arise if one
assumed that the component ${\cal T}^3$ vanished there and thus on
both sides the curves were tangential to the boundary at this position
of space-time. This would be the case for a curve that comes, for
example, from the right, is tangent to the boundary and than goes back
to the right. At the position where the curve is tangential to the
boundary, another curve could flow into this curve from the other side
of the boundary, leading to a branching point. But this situation is
not possible because the tangential vectors of the curves have to
point in the directions of the future light cone. Looking at the
neighboring curves of those considered above, these possible
directions would be inconsistent with the continuity of ${\cal T}^3$
and the fact that the curves do not intersect each other outside the
boundary. Thus there are no branching points at the boundary.

\begin{figure}[htb]
\begin{center}
\leavevmode
\epsfxsize=1.0\linewidth
\epsfbox{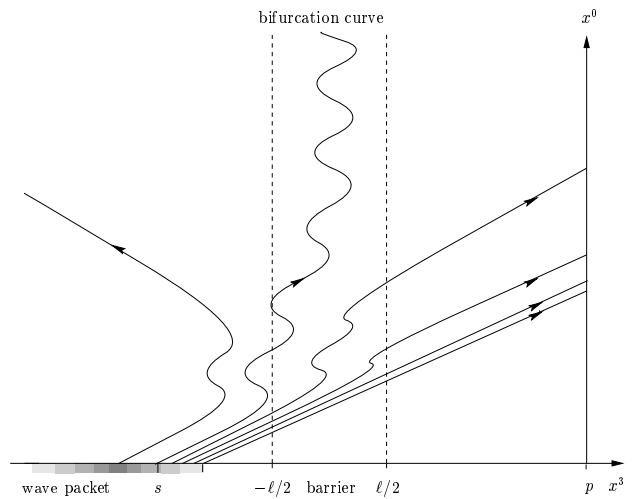}
\caption{A qualitative plot of the energy flow in a $x^0$-$x^3$-plane of the
Minkowskian space-time. The plot intensity of the initial wave packet
corresponds to its energy density.}
\label{fig5}
\end{center}
\end{figure}

So we obtain the qualitative picture for the integral curves of the
energy flow in the space-time shown in Fig.\ \ref{fig5}. The curves
originate in the initial pulse of which the wave front is still to the
left of the barrier at time $x^0=0$. The curve originating in the wave
front of the pulse has a slope of one because the wave front
propagates with the vacuum speed of light. 
The fact that the curves do not
intersect each other allows the initial pulse to be decomposed into
two connected parts from which transmitted and reflected curves,
respectively, originate.  If there is no energy absorbed by the
barrier, the starting position $s$ of the bifurcation curve separating
transmitted and reflected curves is given implicitly by the
transmitted portion of the energy of the initial pulse, i.e.
\begin{equation}
\int\limits_s^{\infty} {\cal T}^0_{|x^0=0} dx^3 = 
\int\limits_0^{\infty} {\cal T}^0_{|x^3=p} dx^0.
\end{equation}
The integration along the time axis can be done at any position 
$x^3=p$ behind the barrier.

Now, by means of the energy flow as shown in Fig.~\ref{fig5}, we want
to explain the existence of solutions with superluminal maxima within
a causal theory. The part of the initial pulse between its wave front
and the starting point $s$ of the bifurcation curve is mapped along
the energy flow on the time axis at $x^3=p$. The closer two
neighboring points of the initial pulse at $x^0=0$ are to the starting
point $s$ of the bifurcation curve, the more the distance between them
is extended by this mapping. This is necessary because of the arrival
of transmitted curves at $x^3=p$ also at arbitrary late times.  Due to
this spreading of the curves the energy density of the transmitted
pulse at $x^3=p$ begins to decrease from a particular time
corresponding to the arrival time of the maximum behind the
barrier. In other words, the transmitted pulse results from a
redistribution of the energy contained in the forward tail between the
front and $s$ of the starting pulse. Thus the maxima of the initial
and transmitted pulse are not causally related. 

The whole picture arises as a consequence of the mathematical claim
that the curves do not intersect each other. These curves themselves,
of course, cannot be observed in any experiment. But we believe 
that they are a suitable tool to get a classical picture of the
mechanism of the tunneling effect. Within this classical 
interpretation the surprising result of our solutions is the almost 
exact reconstruction of a Gaussian wave packet behind the barrier. 
This effect yields a pulse-reshaping \cite{Har}. 

To obtain a more physical point of view one can ask at which time
the transmitted pulse exceeds an arbitrary threshold behind the
barrier. Due to the damping and squeezing of the transmitted pulse 
this happens always at a later time compared to a pulse which crosses 
no barrier. That means for an observer behind the barrier 
that he would not detect the tunneled pulse earlier than the freely 
propagated one, in agreement with causality.

\section{SUMMARY AND CONCLUSION}

In Eq.\ (\ref{finsol}), we have given an exact analytic expression 
in the time domain for the causal Green's function of a model
that describes an ideal case of electromagnetic tunneling. The
structure of this function allows for a reduction to those terms which
describe only the pure tunneling effect without the distortions caused
by the waveguide.  With this reduced Green's function, we calculated
the shape of transmitted wave packets for truncated Gaussians as
initial pulses.  The resulting pulse can be also Gaussian-like. In
agreement with the approximate result of Martin and Landauer
\cite{Mar} and the experiments on microwaves by Enders and Nimtz
\cite{End2}, the time delay of the maximum of the pulse becomes nearly
independent of the thickness for sufficiently long barriers. Furthermore,
the variance of the transmitted packet decreases with increasing
barrier thickness. By examining the properties of the energy
flow, we have found consistency of a
superluminal pulse maximum with the causality of Maxwell's theory.

Within this interpretation the energy of the transmitted pulse can
only originate from a connected part behind the wave front of the
initial wave packet because the integral curves of the energy flow do not
intersect each other. In this sense, the Gaussian shape of the
transmitted pulse can only be interpreted as an amazing interference
effect. Due to the propagation of the wave front with the vacuum speed of
light, it is not possible to obtain a superluminal maximum if the
barrier thickness exceeds the distance between the maximum and the
wave front. Thus, in the case of truncated wave packets, the time
delay of the maximum does not stay independent of the thickness for all
barrier length. In summary, the results of the tunneling experiments
can be obtained from the causal Maxwell theory.

\acknowledgments

The author acknowledges with thanks the stimulating discussions with 
G.~Nimtz, F.W.~Hehl, E.~Mielke and W.~Heitmann, and the critical 
reading of the manuscript by A.~Volmer.

\end{multicols}

\end{document}